\documentclass{article}

\usepackage{arxiv}

\usepackage[utf8]{inputenc}
\usepackage[T1]{fontenc} 
\usepackage{graphicx} 
\usepackage{xcolor}
\usepackage{hyperref}       
\usepackage{url}            
\usepackage{booktabs}       
\usepackage{amsfonts}       
\usepackage{microtype}      
\usepackage{doi}
\usepackage{natbib}
\usepackage{array}
\usepackage{makecell}
\usepackage{changepage}
\usepackage{tabularx}
\usepackage{float}
\usepackage{pifont}

\usepackage{orcidlink}

\title{Dementia Insights: A Context-Based MultiModal Approach}

\author{
Sahar Sinene Mehdoui$^{1}$~\orcidlink{0009-0005-7065-1284}, 
Abdelhamid Bouzid$^{1}$~\orcidlink{0000-0002-2675-9432}, 
Daniel Sierra-Sosa$^{2}$~\orcidlink{0000-0001-5274-8596}, 
Adel Elmaghraby$^{1}$~\orcidlink{0000-0003-1326-0867} \\
$^{1}$ Department of Computer Science and Engineering, University of Louisville\\
\texttt{adel.elmaghraby@louisville.edu} \\
$^{2}$ Department of Electrical Engineering and Computer Science, The Catholic University of America\\
\texttt{sierrasosa@cua.edu}
}
\date{}

\hypersetup{
pdftitle={Dementia Insights: A Context-Based MultiModal Approach},
pdfsubject={cs.LG, q-bio.NC, q-bio.OT},
pdfauthor={Sahar Sinene Mehdoui, Abdelhamid Bouzid, Daniel Sierra-Sosa, Adel Elmaghraby},
pdfkeywords={Dementia, CHAT, speech, text, Deep Learning, Cross Attention, GPT, Bert, CLIP, CLAP, embedding, MLP, In-Context Learning},
}

\begin{document}
\maketitle
\begin{abstract}
	Dementia, a progressive neurodegenerative disorder, affects memory, reasoning, and daily functioning, creating challenges for individuals and healthcare systems. Early detection is crucial for timely interventions that may slow disease progression. Large pre-trained models (LPMs) for text and audio, such as Generative Pre-trained Transformer (GPT), Bidirectional Encoder Representations from Transformers (BERT), and Contrastive Language-Audio Pretraining (CLAP), have shown promise in identifying cognitive impairments. However, existing studies generally rely heavily on expert-annotated datasets and unimodal approaches, limiting robustness and scalability. This study proposes a context-based multimodal method, integrating both text and audio data using the best-performing LPMs in each modality. By incorporating contextual embeddings, our method improves dementia detection performance. Additionally, motivated by the effectiveness of contextual embeddings, we further experimented with a context-based In-Context Learning (ICL) as a complementary technique. Results show that GPT-based embeddings, particularly when fused with CLAP audio features, achieve an F1-score of $83.33\%$, surpassing state-of-the-art dementia detection models. Furthermore, raw text data outperforms expert-annotated datasets, demonstrating that LPMs can extract meaningful linguistic and acoustic patterns without extensive manual labeling. These findings highlight the potential for scalable, non-invasive diagnostic tools that reduce reliance on costly annotations while maintaining high accuracy. By integrating multimodal learning with contextual embeddings, this work lays the foundation for future advancements in personalized dementia detection and cognitive health research.
\end{abstract}


\section{Introduction}
Dementia is an increasing global health challenge, affecting millions of individuals and imposing significant burdens on healthcare systems \cite{who_dementia}. Early diagnosis is critical for facilitating timely interventions that can slow disease progression and improve patient outcomes. Although effective, traditional diagnostic methods such as neuroimaging and biomarker analysis are often costly, invasive, and inaccessible in many settings \cite{el2024navigating, chavez2021diagnosis, zhao2023neuroimaging, lee2017cost}. Consequently, research has increasingly focused on non-invasive, scalable diagnostic methods, particularly those utilizing speech and language analysis to detect cognitive impairments through linguistic and acoustic markers.

Advancements in artificial intelligence (AI) and computational methods have significantly transformed dementia research. Early machine learning models, such as logistic regression and Random Forests, effectively identified cognitive decline using linguistic indicators such as coherence, hesitations, and repetitions \citep{shah2021learning, jahan2024early}. The emergence of deep learning, particularly transformer-based architectures such as BERT \cite{devlin2019bertpretrainingdeepbidirectional} and GPT \cite{radford2019language, brown2020language, adler2024gpt}, has further improved the detection of subtle linguistic patterns associated with cognitive impairments. Specialized adaptations, like AD-BERT \citep{mao2023ad}, have improved early dementia detection through pre-training on Alzheimer’s disease-related datasets. Large language models (LLMs), such as GPT-$4$ \cite{adler2024gpt}, excel in few-shot learning and effectively capture nuanced linguistic expressions linked to cognitive decline \cite{du2024enhancing}

Beyond text-based analyses, audio data provide critical insights into dementia-related speech characteristics by examining features such as pitch, pauses, articulation changes, and prosody. Advanced signal processing techniques, including log-Mel spectrograms and delta coefficients, have proven to be effective in identifying cognitive impairments \cite{meghanani2021exploration}. Vision Transformers (ViTs) \cite{dosovitskiy2021imageworth16x16words} have outperformed traditional CNN-based models in capturing global dependencies from speech signals \citep{ilias2023detecting}, further advancing audio-based dementia detection.

Integrating text and audio data through multimodal approaches holds significant potential for improving classification accuracy. By combining the richness of textual features with complementary speech-based cognitive markers \citep{han2022automatic}, multimodal frameworks such as GP-Net, which \cite{liu2022improving} proposed, have demonstrated superior performance over conventional transformer-based models in dementia detection \citep{liu2022improving}. These integrated approaches underscore the value of harnessing symbolic annotations, linguistic structures, and acoustic properties in providing a more comprehensive assessment of cognitive health.

Despite these advancements, several challenges persist. Many studies rely on structured clinical notes or specialized datasets, thereby limiting their applicability to real-world settings. Moreover, while transcribed speech data include valuable linguistic annotations, such as self-corrections and repetitions, their practical impact on machine learning performance remains underexplored. There is a critical need for dementia detection models that are accurate, scalable, generalizable, and interpretable.

In this paper, we introduce a context-driven multimodal model as our primary contribution, integrating both text and audio features for dementia detection using large pre-trained models. Unlike prior studies, our method explicitly incorporates contextual information across both modalities, enhancing interpretability and robustness. Additionally, recognizing the importance of contextual learning in multimodal analysis, we explore an In-Context Learning (ICL) strategy to evaluate its potential as a complementary technique. Our experiments demonstrate that while ICL shows promise, our proposed framework significantly outperforms alternative methods, making it the core innovation of this study. Previous works, such as \cite{pan2025two}, have leveraged cross-attention mechanisms with self-supervised learning (SSL) models like wav2vec2.0 and transformer-based architectures such as BERT, but these primarily relied on structured datasets and emphasized audio features. This study addresses those limitations by integrating multimodal data with large language models and evaluating both expert-annotated and raw transcriptions to improve scalability and robustness. We leverage the best available pre-trained models in each modality, including CLIP \cite{OpenAI_CLIP}, GPT, and BERT for text embeddings, and CLAP \cite{elizalde2023clap} for audio. Using the Pitt Corpus \cite{macwhinney2011pitt}, we conduct a broad comparative analysis, assessing the efficacy of expert-annotated versus raw datasets. Our findings offer valuable insights into how various text and audio features contribute to dementia detection, underscoring the potential of scalable, non-invasive diagnostic tools.

The structure of this paper is as follows: section \ref{sec:related_work} provides an overview of existing dementia detection methodologies, including text-based, audio-based, and multimodal approaches. Section \ref{sec:preliminaries} describes the \textit{Cookie Theft Test} \cite{cummings2019describing} and the Codes for the Human Analysis of Transcripts (CHAT) \cite{macwhinney2024CHAT}, both of which form the basis of our experimental framework. Section \ref{sec:data_description} describes the dataset utilized in this study. Section \ref{sec:methodologies} presents the methodologies employed in this study. Section \ref{sec:experiments} reports the experimental results and their analysis. Finally, section \ref{sec:conclusion} discusses the implications of our findings and proposes avenues for future research.

\section{Related Work}
\label{sec:related_work}

Dementia detection has advanced substantially with computational methods, offering innovative techniques for identifying cognitive decline \cite{du2024enhancing, mao2023ad, han2022automatic, bouazizi2023dementia, ilias2023detecting, balagopalan2020bert, zhu2021exploring, pan2025two, baevski2020wav2vec, haulcy2021classifying, bt2024performance}. While traditional diagnostic tools such as neuroimaging and biomarkers remain effective, they are often invasive, expensive, and inaccessible in many settings. As a non-invasive, scalable alternative, speech and language analysis has emerged as a promising method for detecting cognitive impairments by examining text and audio changes indicative of dementia. This section reviews related work, categorized by data modality: text-based, audio-based, and multimodal approaches.

\subsection{Text-Based Methods}

Text-based methods have played a pivotal role in dementia detection, primarily focusing on the analysis of clinical notes, structured speech tasks, and spontaneous transcriptions \cite{gauder2021alzheimer, he2016deep, haider2019assessment, hershey2017cnn, pan2025two, du2024enhancing}. Early studies extracted linguistic features such as word frequency, syntactic complexity, and hesitation markers, applying traditional machine learning algorithms such as Random Forests and logistic regression for classification \cite{shah2021learning}.

The introduction of deep learning into dementia research has significantly improved detection models. Transformer-based architectures, such as BERT, BioBERT \cite{lee2020biobert}, and ClinicalBERT \cite{ClinicalBERT_HuggingFace}, have transformed the analysis of unstructured clinical data. Specialized models such as AD-BERT, pre-trained on Alzheimer’s disease-specific datasets, demonstrated competitive performance in forecasting disease progression \cite{mao2023ad}. Large Language Models (LLMs), including GPT and GPT-$4$, have further advanced dementia detection by leveraging few-shot learning capabilities, enabling robust performance with limited labeled data \cite{du2024enhancing, bt2024performance, chen2024performance}.

Despite these advancements, many models treat transcribed speech as standard text, overlooking critical linguistic annotations that provide deeper insights. For instance, \citet{jahan2024early} used the Pitt Corpus, extracting linguistic features such as hesitations, repetitions, and grammatical errors for traditional machine learning models, with Random Forests achieving the best performance. However, their methodology did not fully leverage the structural nuances of CHAT files, potentially missing critical indicators of cognitive decline.

This study addresses these limitations by incorporating both raw text and annotated transcripts from the Pitt Corpus. We focus on symbolic annotations to capture subtle linguistic cues, enhancing model interpretability and predictive performance. Additionally, we integrate transfer learning from pre-trained models to improve generalizability across diverse datasets.

\subsection{Audio-Based Methods}
Audio-based techniques analyze acoustic features such as pitch, speech rate, articulation patterns, and pauses to detect cognitive impairments. Early studies relied on handcrafted features combined with machine learning classifiers such as support vector machines and Random Forests \cite{haulcy2021classifying}.

Recent advances in deep learning have led to the adoption of more sophisticated architectures for analyzing audio data. Vision Transformers (ViTs) have outperformed traditional CNN-based models by capturing global dependencies in speech signals, thereby improving the identification of dementia-related patterns \cite{ilias2023detecting}. The use of audio-based techniques is particularly advantageous for capturing prosodic and paralinguistic cues that are often lost in text transcriptions.

However, these models face challenges related to dataset variability, noise interference, and differences in recording quality, which can limit their generalizability. Our research addresses these issues by incorporating robust pre-trained audio models such as CLAP, allowing for effective feature extraction from noisy, real-world audio samples.

\subsection{MultiModal Approaches}
Combining text, audio, and sometimes visual data has demonstrated significant potential for improving dementia detection accuracy \cite{bouazizi2023dementia, zhu2023evaluating, han2022automatic, lin2024multimodal}. By leveraging complementary strengths across different modalities, these models provide a more holistic assessment of cognitive markers \cite{han2022automatic}.

Structured tasks such as the Cookie Theft Picture Description \cite{cummings2019describing} are widely used to collect multimodal data, enabling joint analysis of linguistic coherence and audio features. Recent studies have demonstrated the effectiveness of integrating BERT-based text embeddings with spectrogram-derived audio features, establishing improved benchmarks for dementia classification \cite{han2022automatic}. Additionally, pre-trained multimodal models such as CLIP \cite{radford2021learning} and BLIP-$2$ \cite{li2023blip} have been explored for their ability to align textual and visual data, enabling zero-shot analysis of cognitive impairments \cite{zhu2023evaluating}.

Despite these advancements, the integration of text and audio data remains underexplored. Many studies focus on a single modality, missing opportunities to leverage the synergy between linguistic and acoustic cues. Moreover, existing multimodal models often require large, labeled datasets, which can constrain their scalability.

Our research addresses these limitations by employing state-of-the-art language models (GPT, BERT) and advanced audio processing techniques (CLAP). We assess the impact of symbolic linguistic annotations and raw transcriptions, evaluating their contributions to dementia detection. Through multimodal embeddings, our framework enhances diagnostic accuracy, robustness, and scalability.

Ultimately, this study highlights the importance of leveraging advanced multimodal AI techniques to improve early dementia detection, paving the way for more accessible, non-invasive diagnostic tools suitable for diverse clinical and community settings.

\section{Preliminaries}
\label{sec:preliminaries}

\subsection{The Cookie Theft Picture Description Task}
\label{subsec:cookie_theft}
\begin{figure}[ht]
    \centering
    \includegraphics[width=0.6\textwidth]{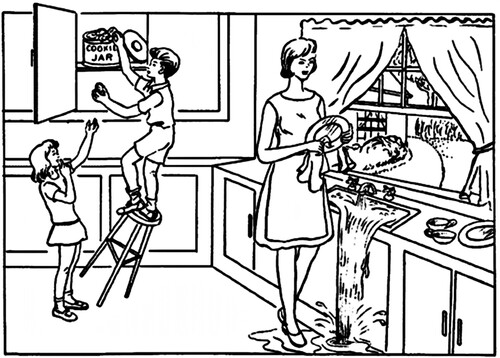} 
    \caption{The Standardized 'Cookie Theft' picture for assessing cognitive abilities through verbal description. Source: Boston Diagnostic Aphasia Examination, available under \href{https://creativecommons.org/licenses/by/4.0/}{Creative Commons Attribution 4.0 International (CC BY 4.0)}. Original image retrieved from \cite{cookie_theft_image}. No changes were made to the original image.}
    \label{fig:cookie_theft_image}
\end{figure}

The Cookie Theft picture description task, as shown in Figure~\ref{fig:cookie_theft_image}, is a well-established cognitive assessment tool used to evaluate multiple cognitive functions. In this task, participants are presented with a standardized image depicting a detailed scene, known as the "Cookie Theft" picture, and are asked to verbally describe their observations in detail.

This task provides insights into several cognitive domains often impaired in dementia. It assesses attention and perception by examining the participant's capacity to explore the image visually, identify key details, and interpret visual elements. Language skills are examined by analyzing clarity, coherence, grammatical accuracy, vocabulary, and fluency in the participant's verbal description. Additionally, executive functioning is assessed through an analysis of the participant’s ability to organize their thoughts, structure a coherent narrative, and logically describe the scene. Memory, particularly working memory, is evaluated based on the participant’s capacity to retain and recall previously mentioned details while tracking multiple elements within the image.

Despite being based on a single standardized image, the task offers a valuable dataset for evaluating a wide range of cognitive functions. This task evaluates not only what participants describe but also how they structure and articulate their responses, providing key insights into dementia-related cognitive impairments. This task was chosen due to its extensive use and demonstrated effectiveness in cognitive assessment.

\subsection{Codes for the Human Analysis of Transcripts (CHAT)}
\label{subsec:cha_analysis}
The Codes for the Human Analysis of Transcripts (CHAT) system, developed for the TalkBank project, provides a structured framework for transcribing and analyzing spoken language \cite{macwhinney2024CHAT}. It uses standardized codes to annotate various linguistic features, including hesitations, grammatical errors, repetitions, self-corrections, and disfluencies. Annotations such as self-corrections, repetitions, aborted speech, and grammatical errors enable a detailed analysis of language patterns that may indicate cognitive decline. These annotations capture nuanced linguistic disruptions often associated with dementia, providing rich data for machine learning models. Table~\ref{tab:cha_differences} highlights the linguistic differences between the Control and Dementia groups.

\begin{table}[ht]
    \centering
    \caption{Summary of Linguistic Differences Between Control and Dementia Groups Based on Visualized CHAT Transcribed Text Data Annotations}
    \label{tab:cha_differences}
    \begin{tabular}{lcc}
        \toprule
        \textbf{Aspect} & \textbf{Control Group} & \textbf{Dementia Group} \\
        \midrule
        Coherence & Clear and structured & Fragmented and incoherent \\
        Disfluencies & Few & Frequent \\
        Grammatical Accuracy & Minimal errors & Frequent errors \\
        Repetition & Minimal & Frequent repetition \\
        Hesitation Sounds & Limited & Frequent \\
        Content Relevance & Focused and relevant & Unrelated or nonsensical \\
        \bottomrule
    \end{tabular}
\end{table}

While CHAT annotations offer in-depth linguistic insights, they present notable challenges. The manual annotation process is time-intensive and requires expertise from linguists and speech therapists, making it resource-demanding and less scalable for large datasets or real-world clinical applications. This reliance on specialized annotators limits its feasibility for widespread use, particularly in settings where rapid, automated assessments are needed. Additionally, the structured nature of CHAT may not fully capture the variability and complexity of natural speech, potentially restricting the generalizability of models trained solely on annotated data.

In this study, we integrate CHAT-annotated transcripts with raw, unannotated text data to evaluate their effectiveness in dementia detection. By comparing models trained on both data types, we aim to determine whether the granular linguistic markers from CHAT significantly enhance model performance compared to raw text processed through advanced language models such as GPT. Furthermore, we combine these linguistic features with audio data to enable multimodal analysis, correlating text-based markers with acoustic cues such as prosody, pitch, and speech rhythm. This approach balances detailed linguistic analysis with scalability, enhancing the potential for practical, non-invasive dementia screening tools.

\section{Dataset Description}
\label{sec:data_description}

\subsection{Overview}
The dataset used in this study is obtained from the Pitt Corpus. This dataset is part of the Alzheimer’s and Related Dementias Study, collected at the University of Pittsburgh School of Medicine \cite{pitt_corpus}. The dataset consists of transcribed text and audio recordings from individuals categorized into two groups: a control group (cognitively healthy individuals) and a dementia group (individuals diagnosed with dementia). For this study, the dataset contains $243$ tests from $99$ control participants and $309$ tests from $194$ dementia participants. Each test comprises a text file containing the transcript of the verbal description and a corresponding \textit{MP3} audio file of the recorded speech.

All participants completed the Cookie Theft Picture Description (Cookie) test. While the other three tests: Verbal Fluency (Fluency), Delayed Story Recall (Recall), and Sentence Repetition (Sentence) have significantly fewer participants in the Dementia group and almost no participants in the Control group. Moreover, only text files are available for these tests, with no corresponding audio recordings. This imbalance in participation and the lack of multimodal data for these tests limited their inclusion in certain analyses, reinforcing the primary role of the Cookie Theft test in this study.

This dataset includes a variable number of participants across different test categories. Table~\ref{tab:dataset_stats} summarizes the distribution of unique participants for each test, separated by control and dementia groups.

\begin{table}[ht]
    \centering
    \caption{Number of Unique Participants in Each Test Category Across Four Speech-Based Cognitive Assessments for Dementia Detection in the Pitt Corpus data}
    \label{tab:dataset_stats}
    \begin{tabular}{lcc}
        \toprule
        \textbf{Test} & \textbf{Control Group} & \textbf{Dementia Group} \\
        \midrule
        Cookie Theft (Cookie) & $99$ & $194$ \\
        Verbal Fluency (Fluency) & $2$ & $163$ \\
        Delayed Story Recall (Recall) & $1$ & $178$ \\
        Sentence Repetition (Sentence) & $1$ & $161$ \\
        \bottomrule
    \end{tabular}
\end{table}

\section{Methodologies}
\label{sec:methodologies}

\subsection{Context-Based In-Context Learning}

In-Context Learning (ICL) is a transformative approach within large pretrained language models, enabling dynamic adaptation to various tasks based solely on contextual information provided during inference. This process eliminates the need for traditional fine-tuning or gradient updates. Instead, ICL utilizes the extensive pre-trained knowledge embedded within these models, using input context such as task instructions, examples, or specific queries to infer task requirements and generate task specific outputs. The core strength of ICL lies in the model's ability to recognize patterns, relationships, and task structures directly from input data.

In this study, we use an ICL-based framework, described in Figure~\ref{fig:ICL_model}, and detailed in Appendix~\ref{subsec:ICL}, to classify speech transcriptions derived from the Cookie Theft Test, categorizing them as belonging to either the dementia or control group. The transcriptions exclusively contain patient speech content, omitting any extraneous metadata to maintain focus on linguistic features relevant to cognitive assessment.

\begin{figure}[H]
    \centering
    \includegraphics[width=0.8\textwidth]{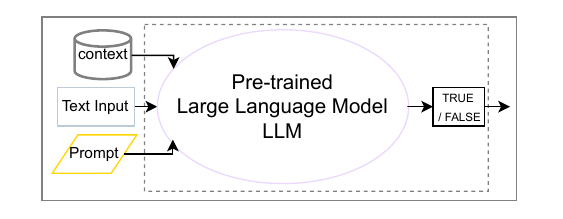}
    \caption{In-Context Learning (ICL) Model Architecture for Dementia Classification Using Structured Prompts with Large Language Models.}
    \label{fig:ICL_model}
\end{figure}

The ICL model architecture integrates a structured prompt designed to effectively guide the model in performing its classification task. This prompt comprises three key components:

\begin{itemize}
\item Task Definition: The model is instructed to act as an expert in the analysis of conversational data for dementia detection, with specific emphasis on linguistic patterns present in Cookie Theft Test descriptions.
\item Classification Criterion: The decision-making process is grounded in the identification of linguistic markers associated with dementia, such as disfluencies, topic drift, reduced syntactic complexity, and other cognitive-linguistic anomalies.
\item Response Format: The model is restricted to generating binary outputs, specifically 'TRUE' to indicate the presence of dementia-related markers and 'FALSE' otherwise.
\end{itemize}

Since LLMs are trained to perform well on general tasks, they struggle to achieve reliable performance on domain-specific data. Thus, providing explicit instructions is essential. An effective way of aiding LLMs in making more accurate decisions is to condition them with a context that includes several examples of data points. Upon receiving this prompt, the pretrained language model processes the transcription, leveraging its in-context reasoning capabilities to generate a classification output.

To assess the efficacy of the ICL approach, we conducted experiments using multiple state-of-the-art large language models.
Additionally, recognizing the impact of context-based ICL approach, we explored a context-based multimodal model, which is the primary contribution of this work and the most effective solution.

\subsection{Context-Based MultiModal Model}

\begin{figure}[h!]
    \centering
    \includegraphics[width=1\textwidth]{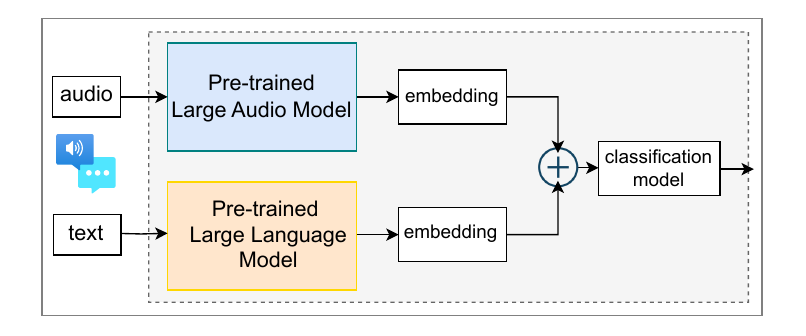} 
    \caption{Schematic Representation of the Multimodal Model Integrating Text and Audio Features.}
    \label{fig:model_architecture}
\end{figure}

The primary contribution of this work is a context-based multimodal model architecture, which improves dementia detection by integrating text and audio data while incorporating contextual information during processing. Our approach leverages advanced pre-trained models, ensuring that both text and speech signals contribute meaningfully to classification. This architecture leverages the complementary strengths of pre-trained Large Language Models (LLMs) and Large Audio Models to capture nuanced cognitive markers. The components of this architecture are illustrated in Figure~\ref{fig:model_architecture} and Appendix~\ref{subsec:multi_config}.

The model processes two primary data modalities: 
\begin{itemize}
    \item Audio Data Processing: The audio recordings are first processed by a pre-trained Large Audio Model, specifically CLAP, which is optimized for capturing intricate acoustic patterns. This model extracts high-dimensional audio embeddings representing features such as prosody, pitch variations, speech rate, and pauses, which serve as critical markers indicative of cognitive decline.
    \item Text Data Processing: Simultaneously, the corresponding speech transcripts are processed by a pre-trained LLM, such as GPT, BERT or CLIP, to generate dense text embeddings. These embeddings encapsulate semantic richness, syntactic complexity, and other linguistic markers relevant to dementia detection.
\end{itemize}

\begin{figure}[ht]
    \centering
    \includegraphics[width=0.8\textwidth]{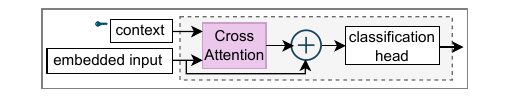}
    \caption{Illustration of Cross-Attention Mechanism within the Multimodal Classification Framework.}
    \label{fig:classification_model}
\end{figure}

The audio and text embeddings are then fused through element-wise addition, creating a unified multimodal representation. This fusion strategy ensures that both modalities contribute equally to the downstream classification task, allowing the model to learn joint representations that capture the synergy between text and audio cues.
The fused embeddings are then fed into a sophisticated classification model, detailed in Figure~\ref{fig:classification_model}, which employs a cross-attention mechanism. This mechanism enables dynamic interaction between the multimodal embeddings and additional contextual data, enhancing classification performance.

The context consists of balanced samples from both the control and dementia groups, consisting of five samples per group, which are dynamically selected during training to: 

\begin{itemize}
\item Expose the model to a diverse range of cognitive patterns, thereby enhancing generalization.
    \item Enable context-aware conditioning, allowing curated contextual inputs during inference to enhance performance without requiring re-training.
    \item Reduce over-reliance on fixed contexts, promoting adaptability to varying input conditions.
\end{itemize}

To reduce information loss often associated with attention mechanisms, we incorporate a residual connection. This reintroduces the original embedded inputs into the cross-attention outputs, ensuring the retention of critical features and enhancing model robustness. The output of the cross-attention layer flows through a fully connected classification head, designed for binary classification (dementia vs. control). This head consists of dense layers with non-linear activations, culminating in a softmax output that generates probabilistic predictions.

\section{Experiments}
\label{sec:experiments}

\subsection{Experiments Set-up}

Informed by the dataset description in Section~\ref{sec:data_description} and the details of the Cookie Theft test outlined in Section~\ref{subsec:cookie_theft}, we designated the Cookie Theft test as the primary task for developing and evaluating methods for early dementia diagnosis. The dataset comprises $99$ control participants who collectively contributed $243$ tests, and $194$ dementia participants with $309$ tests. Each test includes both a CHAT-transcribed text file and an accompanying \textit{MP3} audio recording, providing a rich multimodal dataset.

To ensure robust and unbiased evaluation, we employed a repeated stratified sampling approach similar to 10-fold cross-validation but with a fixed test set proportion. Specifically, for each iteration, we randomly selected $30\%$ of the positive class (dementia group) for the test set and an equal number of samples from the negative class (control group) to maintain balance. The remaining data was used for training. This process was repeated 10 times, ensuring that different subsets of the data were evaluated while maintaining a consistent $30\%$ test set size. We opted for this approach instead of strict $10$-fold cross-validation to ensure a sufficiently large test set, allowing for a more stable evaluation of model performance. To prevent data leakage and ensure fair evaluation, we enforced a strict constraint to ensure that no participant appeared in both the training and testing sets within the same fold. Additionally, each test fold was balanced to contain an equal representation of control and dementia samples, thereby enhancing the generalizability of the model across unseen individuals. More details about the experimental setup, such as the configurations of the neural networks or specifics about the data preprocessing steps can be found in the appendix section.

For both the multimodal architecture and the In-Context Learning (ICL) approach, we maintained consistency in the context size. Each model was provided with a context window comprising $10$ samples, with $5$ from control participants and $5$ from dementia participants. This balanced context selection was instrumental in conditioning the models to learn from diverse cognitive patterns.
\begin{itemize}
    \item Multimodal Architecture: The architecture incorporates two standard cross-attention blocks. Information flows through these blocks using residual connections (skip connections) to mitigate vanishing gradient issues and preserve essential feature representations.
    \item Classification Head: The final classification head consists of two fully connected linear layers. These layers have dimensions of $32$ and $16$, optimized to handle the fused multimodal embeddings and output binary classification results.
\end{itemize}

Given the binary nature of the dementia classification task, we assessed model performance using precision, recall, and F1-score. These metrics provide a comprehensive assessment of the models' capability to accurately detect dementia cases (true positives) while minimizing false positives and false negatives. This evaluation framework ensures a balanced analysis of both sensitivity and specificity, which is critical for diagnostic applications.

\subsection{Context-Based In-Context Learning Experiments}
In this section, we explore the effectiveness of In-Context Learning (ICL) paradigms by employing state-of-the-art large language models (LLMs), including GPT-$4$o \cite{OpenAI_GPT4o}, Gemini Pro \cite{anil2023gemini}, Gemini $1.5$ Pro \cite{team5gemini}, Claude $3$ \cite{Anthropic_Claude}, Claude $3.5$ Sonnet \cite{Anthropic_Claude3.5}, in the context of early dementia detection. The models were provided with contextual information to contextualize dementia-related linguistic patterns, based on expert annotations highlighting specific markers such as language disfluencies, topic drift, and reduced syntactic complexity. The goal was to determine whether these models could leverage such in-context cues to identify dementia-related language patterns without the need for extensive fine-tuning.

The experimental setup consisted of prompting each LLM with raw text data, accompanied by a carefully crafted context that emphasized key dementia indicators. This approach was designed to simulate a zero-shot or few-shot learning scenario in which model performance depends primarily on its capacity to generalize from the provided context. The results, summarized in Table~\ref{tab:model_results_icl}, reveal nuanced differences in performance across the models.

\begin{table}[ht]
    \centering
    \caption{Performance metrics for In Context Learning.}
    \label{tab:model_results_icl}
    \begin{tabular}{>{\centering\arraybackslash}m{4cm} >{\centering\arraybackslash}m{4cm} c c c c}
        \toprule
        \textbf{Text data type} & \textbf{\makecell{Large Language\\ Model}} & \textbf{Precision} & \textbf{Recall} & \textbf{F1 Score}  & \textbf{Accuracy} \\
    \midrule
        \makecell{Raw Text} & GPT-4o & $67.86\%$ & $62.32\%$ & $64.98\%$ & $64.34\%$ \\
    \midrule
        \makecell{Raw Text} & Gemini Pro & $64.37\%$ & $56.16\%$ & $59.99\%$ & $60.00\%$ \\
        \makecell{Raw Text} & Gemini 1.5 Pro & $\textbf{68.38\%}$ & $\textbf{67.93\%}$ & $\textbf{68.16\%}$ & $\textbf{68.16\%}$ \\
    \midrule
        \makecell{Raw Text} & Claude 3 & $72.18\%$ & $63.95\%$ & $67.82\%$ & $68.10\%$ \\
        \makecell{Raw Text} & Claude 3.5 sonnet & $70.40\%$ & $63.59\%$ & $66.83\%$ & $67.12\%$ \\
    \bottomrule
    \end{tabular}
\end{table}

GPT-$4$o achieved an F1 score of $64.98\%$, exhibiting strong performance, albeit trailing the top-performing models. Gemini Pro and Gemini $1.5$ Pro showed moderate gains, with Gemini $1.5$ Pro achieving an F1 score of $68.16\%$, suggesting that improvements in model architecture and training methodologies enhance ICL performance. Claude $3$ outperforms Claude $3.5$ with an F1 score of $67.82\%$ compared with $66.83\%$, indicating strong consistency in detecting dementia-related linguistic features.

Notably, despite the advanced capabilities of these LLMs, their performance in the ICL setting did not surpass the multimodal models that integrated both text and audio data. For instance, the GPT+CLAP multimodal configuration achieved an F1 score of $83.33\%$ when applied to raw text data, significantly outperforming the best ICL models. This disparity highlights the benefits of multimodal inputs, where audio features complement textual information to provide a richer, more comprehensive representation of cognitive markers.

In text-only configurations, GPT's fine-tuned model achieved an F1 score of $81.96\%$, outperforming its ICL counterpart $68.16\%$. This gap suggests that while LLMs are adept at leveraging in-context information, explicit fine-tuning still offers superior performance in specialized tasks like dementia detection. Additionally, the decline in recall across ICL models, particularly for Gemini Pro $56.16\%$, highlights difficulties in reliably identifying all dementia-related instances, likely stemming from the subtlety and variability of linguistic symptoms.

Overall, these findings underscore the potential of ICL as a flexible, resource-efficient approach for dementia detection. However, they further highlight the importance of multimodal data and fine-tuning in achieving the highest levels of diagnostic accuracy. The complementary nature of these methodologies suggests a promising avenue for future research, in which the strengths of ICL and multimodal learning may be effectively integrated to advance cognitive health assessments.

\subsection{Context-Based MultiModal Model Experiments}

In this section, we analyze the performance of our proposed dementia detection models using two reported tables: Table~\ref{tab:model_results1} reports results on an expert-annotated version of the CHAT-transcribed dataset, and Table~\ref{tab:model_results2} reports results on the same dataset without expert annotations (i.e., using raw, unannotated transcriptions). Overall, these experiments evaluate various combinations of pre-trained language models (CLIP, BERT and GPT) alongside an audio model (CLAP) under both single-modal (text-only or audio-only) and multimodal (text+audio) settings.

\begin{table}[ht]
    \caption{Performance of Pre-trained Language and Audio Models on Expert-Annotated Chat Data for Dementia Detection.}
    \label{tab:model_results1}
    \centering
    \begin{tabular}{lcccccc}  
    \toprule
        \textbf{Input Data} & \textbf{Pretrained LLM} & \textbf{\makecell{Pretrained \\ Audio Model}} & \textbf{Precision} & \textbf{Recall} & \textbf{F1 Score} & \textbf{Accuracy} \\
    \midrule
    \makecell{Text Only}  & CLIP & --- & $71.97\%$ & $72.02\%$ & $72.00\%$ & $72.05\%$ \\
        \makecell{Text Only} & BERT & --- & $77.25\%$ & $76.19\%$ & $76.72\%$ & $76.72\%$ \\
        \makecell{Text Only}  & GPT & --- & $79.15\%$ & $78.57\%$ & $78.86\%$ & $78.86\%$ \\
    \midrule
        \makecell{Audio Only} & --- & CLAP & $71.36\%$ & $70.83\%$ & $71.09\%$ & $71.09\%$ \\
        \makecell{Text and Audio} & CLIP & CLAP & $73.17\%$ & $73.21\%$ & $73.19\%$ & $73.22\%$ \\
        \makecell{Text and Audio} & BERT & CLAP & $79.92\%$ & $78.57\%$ & $79.24\%$ & $79.20\%$ \\
        \makecell{Text and Audio} & GPT & CLAP & $\textbf{80.32\%}$ &$ \textbf{80.36\%}$ & $\textbf{80.34\%}$ & $\textbf{80.36\%}$ \\
    \bottomrule
    \end{tabular}
\end{table}

\begin{table}[ht]
    \caption{Performance of Pre-trained Language and Audio Models on Raw (Non-Expert-Annotated) Chat Data for Dementia Detection.}
    \label{tab:model_results2}
    \centering
    \begin{tabular}{lcccccc}  
    \toprule
        \textbf{Input Data} & \textbf{Pretrained LLM} & \textbf{\makecell{Pretrained \\ Audio Model}} & \textbf{Precision} & \textbf{Recall} & \textbf{F1 Score} & \textbf{Accuracy}\\
    \midrule
        \makecell{Text Only} & CLIP & --- & $76.12\%$ & $75.01\%$ & $75.56\%$ & $75.60\%$ \\
        \makecell{Text Only} & BERT & --- & $81.37\%$ & $77.79\%$ & $79.54\%$ & $80.50\%$ \\
        \makecell{Text Only}  & GPT & --- & $83.01\%$ & $80.95\%$ & $81.96\%$ & $81.93\%$ \\
    \midrule
        \makecell{Audio Only} & --- & CLAP & $71.36\%$ & $70.83\%$ & $71.09\%$ & $71.09\%$ \\
    \midrule
        \makecell{Text and Audio} & CLIP & CLAP & $78.96\%$ & $78.57\%$ & $78.76\%$ & $78.66\%$ \\
        \makecell{Text and Audio} & BERT & CLAP & $81.83\%$ & $81.55\%$ & $81.68\%$ & $81.62\%$ \\
        \makecell{Text and Audio} & GPT & CLAP & $\textbf{83.34\%}$ & $\textbf{83.33\%}$ & $\textbf{83.33\%}$ & $\textbf{83.33\%}$\\
    \bottomrule
    \end{tabular}
\end{table}

Across text-only configurations, GPT consistently outperforms BERT and CLIP in both expert-annotated and raw-data scenarios. In the expert-annotated dataset, GPT achieves an F1 score of $78.86\%$, while BERT attains $76.72\%$ and CLIP attains $72.00\%$. For the raw dataset, GPT’s F1 score increases to $81.96\%$, surpassing BERT’s $79.54\%$ and CLIP's $75.56\%$. This finding indicates GPT’s robustness in capturing linguistic cues indicative of dementia, highlighting its capacity to adapt to diverse conversational contexts.

When examining audio-only input using CLAP, performance remains the lowest among all tested configurations, with an F1 score of approximately $71.09\%$. While audio features provide useful information, relying exclusively on audio is insufficient relative to text-based or multimodal approaches. Nonetheless, combining the audio model with text models substantially improves overall performance. The fusion of text (whether CLIP, BERT, or GPT) with CLAP yields higher F1 scores than either text-only or audio-only pipelines. Notably, GPT+CLAP obtains the strongest results, achieving an F1 score of $80.34\%$ on the expert-annotated set and $83.33\%$ on the raw dataset. These trends confirm the complementarity between text and audio features, which together capture a broader spectrum of dementia-related signals.

A noteworthy observation is that the raw (non-expert-annotated) data yields performance that is equal to or even superior to the expert-annotated version. GPT+CLAP, for example, improves from an F1 score of $80.34\%$ in the annotated dataset to $83.33\%$ in the raw dataset. One explanation is that expert annotations may inadvertently introduce confusion by labeling behaviors that can occur in both healthy individuals and those with dementia, thus diminishing model precision. Furthermore, large pre-trained architectures such as GPT and CLAP demonstrate a strong capacity to learn text and audio patterns directly from raw data, reducing the need for detailed manual labeling.

From a practical standpoint, these results are significant because they suggest that expensive and time-consuming expert annotations are not strictly necessary for achieving robust dementia detection. By leveraging pre-trained models capable of extracting nuanced patterns, researchers and clinicians can train effective classifiers even when only raw text is available. Eliminating the requirement for expert annotation also makes the approach more scalable, enabling broader application in clinical and community-based settings. Finally, the consistent gains achieved by combining GPT with CLAP highlight the advantages of multimodal analysis, where language usage and speech properties together yield a more holistic and powerful assessment of cognitive functioning.

\subsection{Comparison and Discussion}

We conducted a comparative analysis of our proposed approach against two state-of-the-art models referenced from recent literature \cite{pan2025two} and \cite{bt2024performance}. The models were selected based on their outstanding performance in dementia detection, as reported in their respective studies. To ensure a fair and rigorous comparison, we re-implemented these models by closely following the methodologies and configurations detailed in their original publications. This process included replicating the architectural frameworks, data preprocessing techniques, and training protocols as described by the authors. Before proceeding with the evaluation, we first ensured that our re-implementations replicated the reported results as closely as possible. This step was crucial to validate the correctness of our re-implementation and ensure that any observed differences in performance were due to methodological advancements rather than inconsistencies in replication. Once we confirmed the fidelity of our implementations, we evaluated all models under the same conditions.

Consistency in experimental design was paramount. Therefore, we adhered strictly to the experimental setup outlined at the beginning of Section \ref{sec:experiments}, which includes the use of the Pitt Corpus and a repeated stratified sampling with a fixed $30\%$ test set test set. This ensured that different subsets of the data were evaluated while maintaining a balanced dataset to mitigate bias. This uniform framework ensured that all models were evaluated under identical conditions, providing a robust basis for comparison.

The comparative analysis juxtaposes these state-of-the-art models with our best-performing configuration. We employed the same evaluation metrics precision, recall, F1-score, and accuracy across all models to maintain consistency and objectivity in performance assessment. Our results, shown in table~\ref{tab:model_comparison}, demonstrate that our approach significantly outperforms the selected state-of-the-art models across all metrics. Specifically, our model achieved better precision, recall, and F1-scores, reflecting its enhanced capability in accurately detecting dementia-related patterns. This performance advantage is attributed to our innovative integration of multimodal data, leveraging both text and audio features through advanced pre-trained models, which capture a broader spectrum of cognitive markers compared to traditional approaches.

\begin{table}[ht]
    \caption{Performance Comparison of State-of-the-Art Models and Proposed Approach.}
    \label{tab:model_comparison}
    \centering
    \begin{tabular}{lcccc}  
    \toprule
        \textbf{Model} & \textbf{Precision} & \textbf{Recall} & \textbf{F1 Score} & \textbf{Accuracy} \\
    \midrule
        \cite{pan2025two} Two-step Attention & $79.38\%$ & $76.79\%$ & $76.87\%$ & $77.01\%$ \\
        \cite{bt2024performance} ICL-bard (Q1+Q2) & $70.15\%$ & $68.75\%$ & $68.92\%$ & $68.99\%$ \\
    \midrule
        Context-Based MultiModal (\textbf{ours}) & $\textbf{83.34\%}$ & $\textbf{83.33\%}$ & $\textbf{83.33\%}$ & $\textbf{83.33\%}$ \\
    \bottomrule
    \end{tabular}
\end{table}

\begin{figure}[h!]
    \centering
    \includegraphics[width=0.5\textwidth]{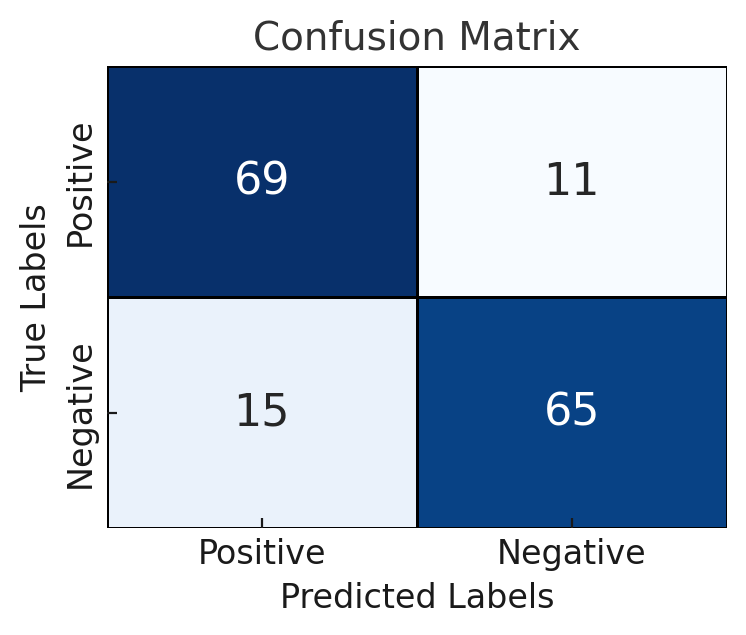} 
    \caption{Confusion matrix for one of the 10 folds.}
    \label{fig:confusion_matrix}
\end{figure}

Figure \ref{fig:confusion_matrix} presents the confusion matrix for one of the randomly selected folds, illustrating the classification performance of the best context-based multimodal configuration (GPT + CLAP). The matrix highlights the distribution of true positives, true negatives, false positives, and false negatives. Notably, the model achieved a high true positive rate, accurately identifying most dementia cases. However, the presence of false positives indicates that some control participants were misclassified as having dementia. 

To better understand the causes of misclassification, we conducted an in-depth analysis of both false positives and false negatives, focusing on text, audio, and multimodal features that may have influenced the model’s predictions. The goal was to identify patterns that could inform future model refinements and enhance classification robustness.

Our analysis revealed that dementia cases misclassified as control often contained structured and coherent narratives, effectively masking subtle cognitive impairments. For instance, clear and detailed descriptions, such as "The mother is drying dishes and the water is cascading onto the floor," made it difficult for the model to detect underlying dementia-related markers. Conversely, control cases misclassified as dementia exhibited frequent hesitations and verbosity, which the model erroneously interpreted as indicators of cognitive decline. 

This limitation likely stems from the pre-trained model’s ability to capture fine-grained linguistic details in embedding representations. Some subtle but crucial features may not have been sufficiently emphasized in the learned embeddings, leading to increased confusion between the two classes. Future work should explore enhancing feature extraction techniques to better distinguish between genuine cognitive impairment and naturally occurring speech variations.

\section{Conclusions}
\label{sec:conclusion}
In conclusion, this study demonstrates the effectiveness of our context-based multimodal approach as the primary innovation in dementia detection. By explicitly incorporating contextual information in both text and audio processing, we achieved state-of-the-art results while improving model robustness and interpretability. Additionally, inspired by the success of context in multimodal learning, we explored a context-based In-Context Learning (ICL) approach, which, while promising, did not surpass the multimodal framework. Our findings highlight the superiority of context-aware multimodal models, paving the way for more reliable and scalable dementia detection methods. GPT-based embeddings consistently outperformed BERT and CLIP, showing superior precision, recall, and F1 scores, likely due to their ability to capture intricate linguistic nuances. The integration of audio features via the CLAP model further improved accuracy, with raw, unannotated data outperforming expert-annotated versions. This suggests that large pre-trained models can extract meaningful patterns without extensive manual annotation, enhancing scalability and real-world applicability. These findings are supported by the comparative performance results presented in Table~\ref{tab:model_comparison}, where our proposed model achieved an F1-score of $83.33\%$ and an accuracy of $83.33\%$, outperforming state-of-the-art models such as the Two-step Attention model \cite{pan2025two} and the ICL-bard (Q1+Q2) model \cite{bt2024performance}.

Furthermore, this work establishes a foundational framework for future research aimed at providing more personalized insights and feedback to dementia patients. By enabling the development of diagnostic tools that go beyond detection, future studies can focus on delivering actionable cognitive health insights tailored to individual patients. This will not only support early diagnosis but also contribute to personalized intervention strategies, empowering clinicians with deeper cognitive analyses. Ultimately, this approach has the potential to improve patient outcomes, foster more precise monitoring of disease progression, and offer valuable resources for both experts and researchers to better understand the complex dynamics of dementia.

\section*{Funding}
This research received no external funding.

\section*{Ethics Statement}
This study did not involve direct interaction with human participants or the creation of new patient data. The data used in this study were obtained from the publicly available Pitt Corpus dataset, which was collected and shared by the Alzheimer and Related Dementias Study at the University of Pittsburgh School of Medicine.

\section*{Data Availability}
The data used in this study are publicly available as part of the Pitt Corpus dataset. The dataset can be accessed through the DementiaBank project at \url{https://dementia.talkbank.org/access/English/Pitt.html}, subject to the approval of the data administrators. This study did not generate any new data.

\section*{Conflicts of Interest}
The authors declare no conflicts of interest.

\section*{Abbreviations}
The following abbreviations are used in this manuscript:

\begin{tabular}{@{}ll}
BERT & Bidirectional Encoder Representations from Transformers \\
GPT & Generative Pre-trained Transformer \\
CLIP & Contrastive Language-Image Pretraining \\
CLAP & Contrastive Language-Audio Pretraining \\
CHAT & Codes for the Human Analysis of Transcripts \\
MLP & Multilayer Perceptron \\
F1 & Harmonic Mean of Precision and Recall \\
\end{tabular}

\appendix

\section{Context-based In-Context Learning (ICL) Setup}
\label{subsec:ICL}

To ensure a controlled evaluation, all models received the same structured input.

\textbf{Note:} This appendix does not contain real patient data. Instead, we present \textit{generic} examples to illustrate the input structure used. All transcript data in this study were anonymized and handled in accordance with ethical research standards.

The content provided to the language models is as follows:

\noindent\fbox{%
\parbox{0.98\textwidth}{%
\textbf{You are a neurologist specializing in speech pattern analysis for dementia detection.} \\
Your task is to analyze patient transcripts from the Cookie Theft picture description test and classify them as either showing signs of dementia (\texttt{TRUE}) or not (\texttt{FALSE}). \\
Follow the classification criteria strictly, and answer using only \texttt{TRUE} or \texttt{FALSE}.

\textbf{Task:} You will be given a transcript of a patient describing the Cookie Theft picture from the Boston Diagnostic Aphasia Examination. Your task is to classify the speech as either showing signs of dementia (\texttt{TRUE}) or not showing signs of dementia (\texttt{FALSE}) based only on the transcript.

\textbf{Key Considerations:}

\ding{51} Normal Aging May Include:
\begin{itemize}
    \item Occasional word-finding difficulty but coherent speech.
    \item Slight redundancy or mild repetition of ideas.
    \item Slower speech but still structured and logical.
\end{itemize}

\ding{55} Dementia May Include:
\begin{itemize}
    \item Disorganized or fragmented speech (sentences missing logical connections).
    \item Frequent grammatical errors (incorrect verb tense, missing words).
    \item Noticeable forgetting of prior statements (repeating an idea separately).
    \item Excessive hesitation sounds ("uh," "um," "you know").
    \item Failure to describe major elements of the picture (key omissions).
    \item Tangential or unrelated comments (mentioning details that do not exist in the picture).
\end{itemize}

\textbf{Expected Features of a Normal Response:}
\begin{itemize}
    \item Mentions key elements of the picture, such as:
    \begin{itemize}
        \item The mother washing or drying dishes.
        \item The boy reaching for cookies.
        \item The stool falling.
        \item The girl near the boy.
        \item Water overflowing from the sink.
    \end{itemize}
\end{itemize}

\textbf{Response Format:}  
Based only on the transcript provided, answer with one word:
\begin{itemize}
    \item \texttt{TRUE} → Signs of dementia are present.
    \item \texttt{FALSE} → No signs of dementia.
\end{itemize}

TRANSCRIPT: <add participant text here>
}}

\bigskip

The following are \textit{generic} transcripts created for illustration purposes. They are not real patient data.

\begin{table}[H]
\caption{Example input-output pairs from ICL experiments.\label{tab:icl_examples}}
\centering
\begin{tabular}{p{0.5\textwidth} c c}
\toprule
\textbf{Input Transcript} & \textbf{Expected Output} & \textbf{Model Response} \\
\midrule
"A boy is on a stool reaching for cookies. The mother is washing dishes. Water is overflowing." 
& \texttt{FALSE} & \texttt{FALSE} \\
\midrule
"The, uh, kid is, um, doing something? And, um, the lady is, uh, water? I don't know." 
& \texttt{TRUE} & \texttt{TRUE} \\
\bottomrule
\end{tabular}
\end{table}

\section{Context-Based MultiModal Model Implementation Details}
\label{subsec:multi_config}

\subsection{Data Preprocessing}
We adopted the preprocessing pipelines of the respective pre-trained models to ensure consistency. Specifically, for the annotated text data, we ensured that only the patient’s speech and the expert annotations were retained, excluding any extraneous information.

\subsection{Model Architecture Configuration}
\begin{itemize}
    \item \textbf{Hidden Dimension Alignment:} Audio embeddings are projected using a linear layer to match the hidden dimension of text embeddings.
    \item \textbf{Cross-Attention Mechanism:} 
    \begin{itemize}
        \item \textbf{Number of Attention Heads:} 8
        \item \textbf{Hidden Dimensions:} $768$ for BERT, $1536$ for GPT
    \end{itemize}
    \item \textbf{Classification Layers:} 
    \begin{itemize}
        \item Fully Connected Layers: $[32, 16]$
    \end{itemize}
\end{itemize}

\subsection{Training Configuration}
\begin{itemize}
    \item \textbf{Loss Function:} Cross-Entropy Loss
    \item \textbf{Optimizer:} AdamW
    \item \textbf{Batch Size:} $4$
    \item \textbf{Number of Epochs:} $200$
    \item \textbf{Learning Rate:} $0.0001$
    \item \textbf{Context Size:} $10$
\end{itemize}

\bibliographystyle{unsrtnat}
\bibliography{main}
\end{document}